\documentclass[aps,pra,twocolumn,showpacs,groupedaddress,amsmath,amssymb]{revtex4}
\usepackage{graphicx}
\usepackage{color}
\usepackage{bm}
\usepackage{float}
\usepackage{mathtools}
\usepackage{nccmath}

\newcommand{\fa}{\mathfrak{a}}

\newcommand{\bK}{\mathbf{K}}

\newcommand{\bM}{\mathbf{M}}
\newcommand{\bN}{\mathbf{N}}
\newcommand{\bP}{\mathbf{P}}
\newcommand{\bQ}{\mathbf{Q}}

\newcommand{\cE}{\mathcal{E}}

\newcommand{\cG}{\mathcal{G}}

\newcommand{\cM}{\mathcal{M}}

\newcommand{\cS}{\mathcal{S}}

\newcommand{\be}{\begin{equation}}
	\newcommand{\ee}{\end{equation}}
\newcommand{\bea}{\begin{eqnarray}}
	\newcommand{\eea}{\end{eqnarray}}

\newcommand{\ed}{\end{document}}

\newcommand{\bi}{\begin{itemize}}
\newcommand{\ei}{\end{itemize}}

\newcommand{\bce}{\begin{center}}
\newcommand{\ece}{\end{center}}

\begin{document}

\title{Asymmetric Localization by Second Harmonic Generation}

\author{H. Ghaemi-Dizicheh$^1$, A. Targholizadeh$^1$, B. Feng $^2$, H. Ramezani$^1$}
\email {hamidreza.ramezani@utrgv.edu}
\affiliation{$^1$Department of Physics and Astronomy, University of Texas Rio Grande Valley, Edinburg, Texas 78539, USA\\$^2$ School of Mathematical and Statistical Sciences, University of Texas Rio Grande Valley, Edinburg, TX 78539, USA}

\begin{abstract}
	We introduce a nonlinear photonic system that enables asymmetric localization and unidirectional transfer of an electromagnetic wave through the second harmonic generation process. Our proposed scattering setup consists of a non-centrosymmetric nonlinear slab with nonlinear susceptibility $\chi^{(2)}$ placed to the left of a one-dimensional periodic linear photonic crystal with an embedded defect. We engineered the linear lattice to allow the localization of a selected frequency $2\omega_\star$ while frequency $\omega_\star$ is in the gap. Thus in our proposed scattering setup, a left-incident coherent transverse electric wave with frequency $\omega_\star$ partially converts to frequency $2\omega_\star$ and becomes localized at the defect layer while the unconverted remaining field with frequency $\omega_\star$ exponentially decays throughout the lattice and gets reflected. For a right-incident wave with frequency $\omega_\star$ there won't be any frequency conversion and the incident wave gets fully reflected. Our proposed structure will find application in designing new optical components such as optical sensors, switches, transistors, and logic elements.
	
\end{abstract}
\maketitle
\section{Introduction} The method of generating a localized mode in periodic structures has paved its footprints in some photonics systems. One can achieve this localization by breaking the translation symmetry through embedding a defect in a periodic lattice \cite{joannopoulos1995photonic,noda2000trapping,akahane2003high,song2005ultra,navadeh2020localized, 2020}. Photonic crystal lasers \cite{painter1999two,park2004electrically,ellis2011ultralow}, strain field traps \cite{sievers1988intrinsic}, strong photon localization \cite{john1987strong}, and mode selection \cite{poli2015selective} are instances for applications of defect mode in periodic photonic systems. \\  Because of the time-reversal symmetry, the photon confinement happens regardless of the direction of the incident electromagnetic wave. In recent years, the applications of asymmetric photonic transport have drawn attention in optical systems \cite{2010,lira2012electrically,2013,wang2013optical,2014, fujita2000waveguide}. To create a system with nonreciprocal light propagation characteristics, we can apply some techniques such as magnetic biasing \cite{jalas2013and,saleh2019fundamentals,hr2014}, and spatiotemporally modulating index of refraction \cite{yu2009complete}. In the last method, the frequency and wavevector of the photon shift simultaneously during the photonic transition process. One can embed a defect in spatiotemporally periodic modulated photonic lattice to localize photons in a non-reciprocal manner \cite{ramezani2018nonreciprocal}. While spatiotemporal modulation is a powerful method to achieve unidirectional localization, achieving such modulation in practice, specifically in a high-frequency regime, is arduous. Consequently, it is imperative to propose a new photonic structure capable of localizing photons asymmetrically.\\
This paper provides a technique of asymmetric photon localization by exploiting second-harmonic generation (SHG). The process of SHG is the well-known observation in nonlinear optics where an electromagnetic wave, called a fundamental wave (FW), interacts with the nonlinear material and generates a new wave with twice the frequency of initial light. The generated wave is referred to as the second harmonic wave (SHW). To achieve the asymmetric localization, we introduce an optical setup consisting of a nonlinear slab located to the left side of a linear periodic lattice. By embedding an engineered defect in the linear lattice, we show that the generated SHW becomes localized with a finite transmission while the remaining unconverted FW gets reflected by the bandgap. In the opposite direction, the photon in the same frequency as the fundamental wave gets totally reflected.\\
We employ the transfer matrix approach to study our scattering system. This approach provides us with a theoretical understanding of the idea investigated in this paper for asymmetric localization. The transfer matrix approach for a second harmonic generation as a method has been developed in optical systems to overcome low nonlinear conversion efficiency \cite{li2007second,ren2010enhanced,li2015application}. We apply a similar approach in a different context to deal with our scattering problem.  Indeed, we follow a formalism that is introduced in refs. \cite{mostafazadeh2018blowing,mostafazadeh2019nonlinear} for a nonlinear scattering process. The main characteristic of the transfer matrix for nonlinear scattering potential is that its entries depend on the amplitudes of incoming waves. This property in some nonlinear scattering problems makes it complicated to find the nonlinear transfer matrix. However, despite its severity, the main motivation for using a nonlinear transfer matrix is its composition rule. 

For the SHG process, the fundamental and generated electromagnetic fields satisfy a system of nonlinear coupled Helmholtz equation \cite{boyd2020nonlinear}. The solution of these equations in a slowly varying envelope approximation is given in terms of the Jacobi elliptic functions. It demands cumbersome calculations to find the exact form of the entries of the nonlinear transfer matrix. Instead, we subject our method to a limitation in the nonlinear process for the second-harmonic generation to overcome this difficulty. In this limitation, we suppose that the energy conversion to the second harmonic is very low such that the fundamental wave remains essentially undepleted. This process is known as a non-depletion approximation in a second harmonic generation \cite{2019}. To show that our approximation method gives a qualitatively correct result in the last section, we use the finite-element method to demonstrate asymmetric photon localization in a system with similar composition.
\section{Scattering Setup}
\begin{figure}
		\includegraphics[scale=0.65]{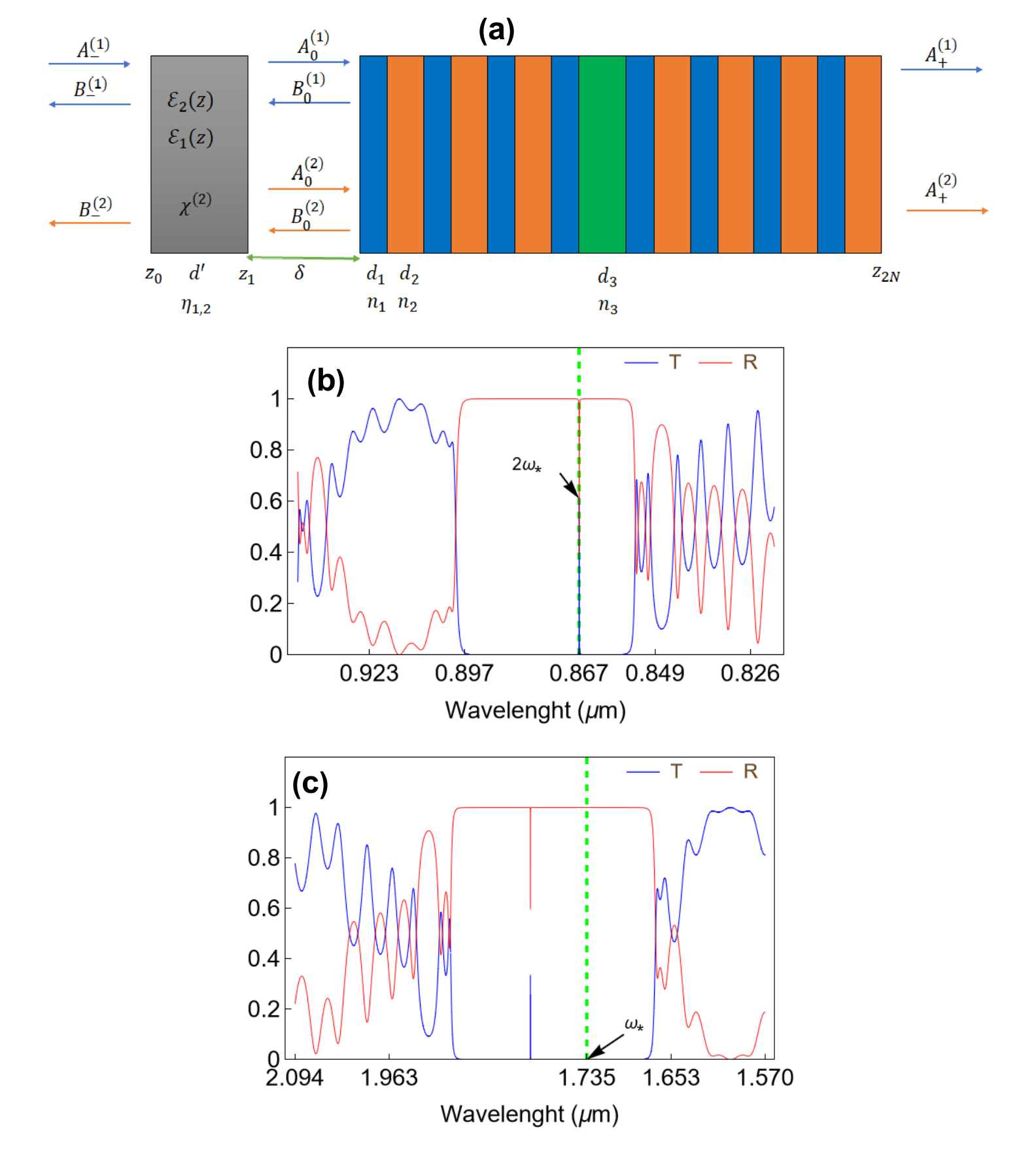}
		\caption{{(a) Schematic diagram of scattering setup consists of a nonlinear material (gray) and one-dimensional defective photonic crystal. The defect is depicted in green. In this diagram, to embed a defect layer, the linear lattice consists of eight segments in which the second section of segment four is manipulated. (b,c) Transmission $ T:=\vert t\vert^2 $ and Reflection $ R:=\vert r \vert^2 $ spectrum of the defective linear photonic crystal in the ranges of $ \lambda $=0.924-0.821 $ \mu m $ and $ \lambda $=2.094–1.570 $ \mu m $. The arrows indicate the transmition amplitude for the FW (with frequency $ \omega_\star $) and SHW (with frequency $ 2\omega_\star $).}} \label{fig:setup}
\end{figure}

To facilitate the semi-analytical approach, let's consider a photonic scattering setup as depicted in Fig.\ref{fig:setup} includes a nonlinear material with nonlinear susceptibility $ \chi^{(2)} $ located to the left of a one-dimensional photonic crystal structure with a distance $ \delta $. The linear lattice is made of $N$ segments of length $ d $, and each segment consists of two homogenous slabs. The thickness and the refractive index of section I is respectively $ d_1 $ and $ n_1(\omega) $, while those for section II is $ d_2 $ and $ n_2(\omega) $, then we have $ d=d_1+d_2 $. Here we consider that, in general, a medium's refractive index depends on the frequency $(\omega) $ of passing light. We can embed a defect by breaking the translation symmetry in the periodic linear lattice. To do this, we manipulate segment $ N_i $ by changing one of its slabs thickness or substitute it with dissimilar material whose refractive index is different from the other part of the lattice.\\
In Fig.\ref{fig:setup}, the gray slab is demonstrating the nonlinear slab that generates the SHW. On the right side, we show the defect layer of the linear periodic structure in green and $ d_3 $ and $ n_3(\omega) $ respectively, stand for its thickness and refractive index.\\
Consider a normally left incident electromagnetic wave with frequency $ \omega=\omega_1 $ hits the system in which its electric field is given by
\be
\vec{E}(\vec{r},t)=\mathcal{E}(z)e^{ik_1x}e^{-i\omega_1 t}\hat{e}_y,
\ee
where $ k_1:=\omega_1/c $ is the wavenumber, $ c $ is the speed of light in vacuum, $ \mathcal{E}(z) $ is the complex
amplitude of the electric field, and $ \hat{e}_j $ is the unit vector pointing
along the $ j $ axis for $ j=x,y,z $. The propagation of the fundamental wave in the nonlinear slab $ \mathcal{S}_1 $ excites the second nonlinear polarization $ \bP_{NL}^{(2)}=\epsilon_0 \chi^{(2)}\vec{E}\cdot\vec{E} $. The induced polarization by the nonlinear medium acts as a source and creates second harmonic field with frequency $ \omega_2:=2 \omega_1 $. The whole nonlinear process can be described by the following nonlinear couple equation \cite{boyd2020nonlinear}:
\begin{eqnarray}
	&&\mathcal{E}''_1+\eta_1^{2}k_1^2\mathcal{E}_1=-k_1^{2}\chi^{(2)}\mathcal{E}^{*}_1\mathcal{E}_2,\label{nceq1}\\
	&&\mathcal{E}''_2+\eta_2^{2}k_2^2\mathcal{E}_2=-k_2^{2}\chi^{(2)}\mathcal{E}_1^{2}.\label{nceq2}
\end{eqnarray}
Here, $ k_2:=\omega_2/c $ is the wavenumber of the SHW and $ \eta_1:=n(\omega_1)$ and $ \eta_2:=n(\omega_2)$ are respectively the refractive index of the nonlinear medium for the fundamental $ \mathcal{E}_1 $ and second harmonic wave $ \mathcal{E}_2 $. In the case of the undepleted fundamental wave, the right side of equation \ref{nceq1} becomes infinitesimal in comparison with the left side, and we can neglect it. Then Eq. \ref{nceq1} admits the following linear solution
\be
\mathcal{E}_1=C^{(1)}e^{ik_1\eta_1z}+D^{(1)}e^{-ik_1\eta_1z}.
\ee
Here, $C^{(1)}$ and $D^{(1)}$ are the complex-valued plane wave coefficients of FW and, respectively, denote forward and backward propagating waves. Then by substituting $ \mathcal{E}_1 $ from the above equation through \ref{nceq2}, Eq. \ref{nceq1} transforms to a nonhomogeneous wave equation such that the right-hand side acts as a source for the SHW induced by FW through the second-order coefficient $\chi^{(2)}$ of the nonlinear medium. We can express the solution of \ref{nceq2} as follows
\bea
&\cE_2(z)=&C^{(2)}e^{ik_2\eta_{2} z}+D^{(2)}e^{ik_2\eta_2z}+\nonumber\\
&&\chi^{(2)}\cG_1[(C^{(1)})^2e^{2ik_1\eta_1z}+(D^{(1)})^2e^{-2ik_1\eta_1z}]\nonumber\\
&&+\chi^{(2)}\cG_2 C^{(1)}D^{(1)},\label{nondepletion sol.}
\eea
where $ \cG_1:=-k_2^2/(k_2^2\eta_2^2-4k_1^2\eta_1^2) $, $ \cG_2:=-1/\eta_2^2 $ and, similar to the pumping wave, $C^{(2)}$ and $D^{(2)}$ are the complex-valued plane wave coefficients of SHW.\\ By adding a photonic crystal on the right side of slab $ \mathcal{S}_1 $, we construct a scattering setup for both fundamental and second harmonic waves. In other word, we wish to construct a scattering solution in the form
\bea
\mathcal{E}_{1,2}(z):=\left\{\!\!\begin{array}{ccc}
	A_{-}^{(1,2)} e^{ik_{1,2} z}+B_{-}^{(1,2)} e^{ik_{1,2}z} & z<z_0,\\[6pt]
	\cE_{1,2}(z)  & z\in[z_0,z_1],\\[6pt]
	A_{j}^{(1,2)} e^{ik_{1,2}n_{j}z}+B_{j}^{(1,2)} e^{-ik_{1,2}n_{j}z}  & z\in[z_j,z_{j+1}],\\[6pt]
	A_{+}^{(1,2)} e^{ik_{1,2}z} & z>z_{2N},
\end{array}\right.\nonumber\\
\label{sol21}
\eea
where $ j:=1,\dots,2N $ labels sections in the multilayer film $\mathcal{S}_2$.\\
We aim to show that for an incident wave with specific frequency $\omega_1:=\omega_\star$, our scattering system transfers the generated wave with frequency $ 2\omega_\star $ and bans the incident fundamental wave from the left, i.e., $A_+^{(1)}=0$. For the right incoming wave, our system completely acts as a mirror and fully reflects the fundamental wave with frequency $ \omega_\star $.\\
We theoretically analyze our system by applying the transfer matrix approach. In this approach, we consider the scattering process of the left/right incident wave in the following steps:\\
a. The fundamental wave propagating in the nonlinear medium induces the nonlinear polarization in which it radiates SHW. In general and without considering non-depletion fundamental wave, the propagation of the initial fundamental wave and the second harmonic wave can be characterized by the nonlinear transfer matrix $ \bM_{N}^{(1,2)} $. The entries of the nonlinear transfer matrix for both FW and SHW can be given in terms of the solutions of \ref{nceq1} and \ref{nceq2} for $ \cE_{1,2} $. The main characteristic of the nonlinear transfer matrix is that its entries depends on the incident amplitudes $ A_-^{(1,2)} $ and $ B_-^{(1,2)} $ and they are not unique \cite{mostafazadeh2019nonlinear}. However the undepleted wave, the nonlinear transfer matrix for the fundamental wave reduces to the linear one.\\
b. The FW and SHW created inside the nonlinear medium propagate through the linear periodic structure. The scattering of both waves can be given by the linear transfer matrix $ \bM_L(\omega) $. The combination of these two steps can be expressed by a single transfer matrix given by
\be
\bM^{(1,2)}=\bM_L\cdot\bM_{N}^{(1,2)}.
\label{compose2}
\ee
In particular, for $ n_1(\omega)=n_0=1 $, the linear transfer matrix $\bM_L$ is uniquely determined by the reflection and transmission amplitudes of the linear crystal $ \cS_2 $. They are given by the following relation:
\begin{align}
	&[\bM_L]_{11}=t-\dfrac{r_lr_r}{t},&&[\bM_L]_{12}=\dfrac{r_r}{t},\nonumber\\
	&[\bM_L]_{21}=-\dfrac{r_l}{t},&&[\bM_L]_{12}=\frac{1}{t},\label{Linear T and R}
\end{align}
where $ t:=t(\omega) $ is the transmittion coefficient, and $ r_l:=r_l(\omega) $ ($r_r:=r_r(\omega)   $) is left (right) reflection coefficient. According to Eqs. \ref{compose2} and \ref{Linear T and R} and in light of the scattering solution \ref{sol21}, the transmitted and reflected wave of the generated wave is given by the following relation
\bea
&&\left(\begin{array}{cc}
	A_{+}^{(2)}   \\
	0
\end{array}\right)=\bM^{(2)}\left(\begin{array}{cc}
	A_{-}^{(2)}    \\
	B_{-}^{(2)}
\end{array}\right)=\bM_L^{(2)}\cdot\bM_{N}^{(2)}\left(\begin{array}{cc}
	A_{-}^{(2)}    \\
	B_{-}^{(2)}
\end{array}\right)\nonumber\\
&&=\bM_L^{(2)}\left(\begin{array}{cc}
	A_{0}^{(2)}    \\
	B_{0}^{(2)}
\end{array}\right)=\dfrac{1}{t}\left(\begin{array}{cc}
	[ t^{2}-r_lr_r] A_{0}^{(2)}+r_rB_{0}^{(2)}    \\
	-r_lA_{0}^{(2)}+B_{0}^{(2)}
\end{array}\right)
.\nonumber\\
\eea
This in turns implies
\begin{align}
	&B_{0}^{(2)}-r_l(2\omega)A_{0}^{(2)}=0,&A_+^{(2)}=t(2\omega) A_{0}^{(2)}.\label{SHG scattering solution}
\end{align}
Following the similar steps for the FW we have
\begin{align}
	&B_{0}^{(1)}-r_l(\omega)A_{0}^{(1)}=0,&A_+^{(1)}=t(\omega)A_{0}^{(1)}.\label{FW scattering solution}
\end{align} 
In the above equation, the intermediate amplitudes $ A_0^{(1,2)}$ and $ B_0^{(1,2)}$ are complex-valued functions of incident amplitudes. Eqs. \ref{SHG scattering solution} and \ref{FW scattering solution} form a system of complex-valued equations where one can solve it to get the scattering amplitudes $ B_-^{(1,2)} $ and $ A_+^{(1,2)} $.\\
For the right-incident wave, the scattering solution is given by \ref{sol21} for $ z\in(z_0,z_{2N}) $ and for elsewhere, we have
\bea
\mathcal{E}(z):=\left\{\!\!\begin{array}{ccc}
	B_{-}^{(1,2)} e^{-ik_{1,2}z},  & z<z_0,\\[6pt]
	A_{+}^{(1,2)} e^{ik_{1,2}z}+B_{+}^{(1,2)} e^{-ik_{1,2}z} & z>z_{2N},
\end{array}\right.\nonumber\\
\label{sol12}
\eea
The scattering amplitudes of the system for the right-incoming wave can be given by the following transfer matrix
\be
\bM^{(1,2)}=\bM_{N}^{(1,2)}\cdot\bM_L^{-1},
\label{compose2right}
\ee
where $ \bM_L^{-1} $ is the inverse of the linear transfer matrix and $ \bM_{N}^{(1,2)} $ is the nonlinear transfer matrix that its entries depends on $ A_+^{(1,2)} $ and $ B_+^{(1,2)} $. By applying the above transfer matrix, we find
\begin{align}
	&B_-^{(j)}=\dfrac{\det \bM_{N}^{(j)}}{[\bM_{N}^{(j)}]_{11}}B_0^{(j)},&A_0^{(j)}=-\dfrac{[\bM_N^{(j)}]_{12}}{[\bM_N^{(j)}]_{11}}B_0^{(j)},
\end{align}
where $ j=1,2 $ and
\bea
&&B_0^{(j)}=\dfrac{r_l(\omega_j)}{t(\omega_j)}A_+^{(j)}+[t(\omega_j)-\dfrac{r_l(\omega_j)r_r(\omega_j)}{t(\omega_j)}]B_+^{(j)},\nonumber\\
\label{R Scattering Eq1}\\
&&A_0^{(j)}=\dfrac{1}{t(\omega_j)}A_+^{(j)}-\dfrac{r_l(\omega_j)}{t(\omega_j)}B_+^{(j)}.\label{R Scattering Eq2}
\eea
In general, the entries of the nonlinear transfer matrix $ \bM_{N} $ can be given in terms of the exact solutions of Eqs. \ref{nceq1} and \ref{nceq2} \cite{boyd2020nonlinear}. In light of them, the system of equations \ref{SHG scattering solution} and \ref{FW scattering solution} consists of Jacobi elliptic functions where scattering amplitudes appear in the elliptic integral of the first kind. Therefore, finding the analytic scattering solution of the second harmonic generation is possible but complicated.\\
\begin{figure}
	\begin{center}
		\includegraphics[scale=0.25]{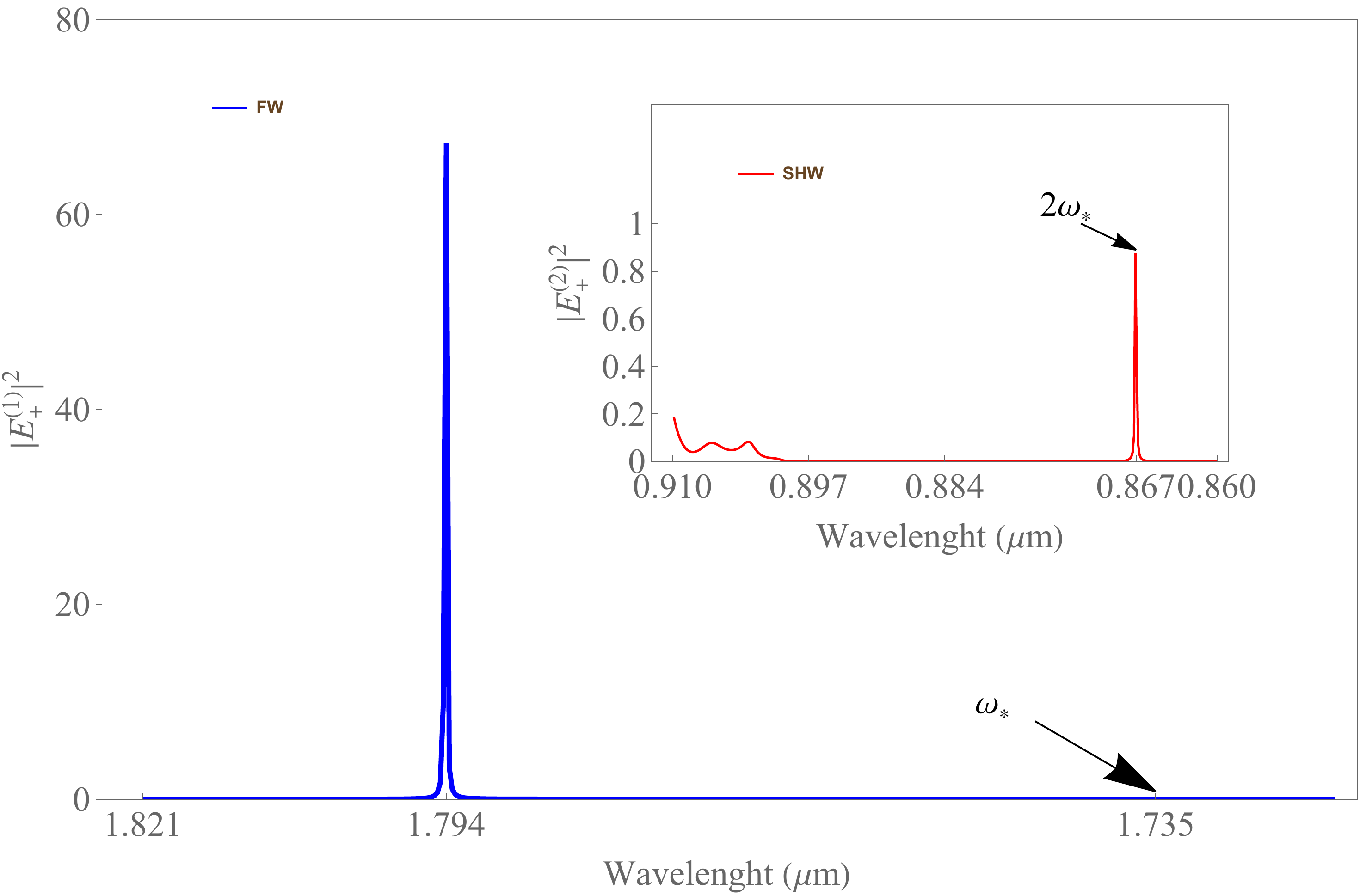}
		\caption{{ Transmitted intensity $ \vert E_+^{(1,2)}\vert^2=\vert A_+^{(1,2)}\vert^2 $ of the FW (blue line) and SHW (red line). The arrows indicate the value of transmitted intensity at $ \lambda_\star $ and $ 2\lambda_\star $.}}  \label{fig:transmittion}
	\end{center}
\end{figure}
Under the non-depletion limit, the second harmonic generation admits a manageable solution \ref{nondepletion sol.} which makes it feasible to find the entries of the nonlinear transfer matrix. Given the non-depletion regime, the nonlinear transfer matrix of the fundamental wave $ \bM_N^{(1)} $ reduces to the linear one. In the appendix, we introduce the nonlinear transfer matrix for the SHW that illuminates the system from the left. Here, we suppose that there is no left incident wave with the same frequency as the second harmonic, i.e., $ A_-^{(2)}=0 $ and the second harmonic reflection emerges through the reflection from the surfaces.\\ 
\begin{figure}
	\begin{center}
		\includegraphics[scale=.35]{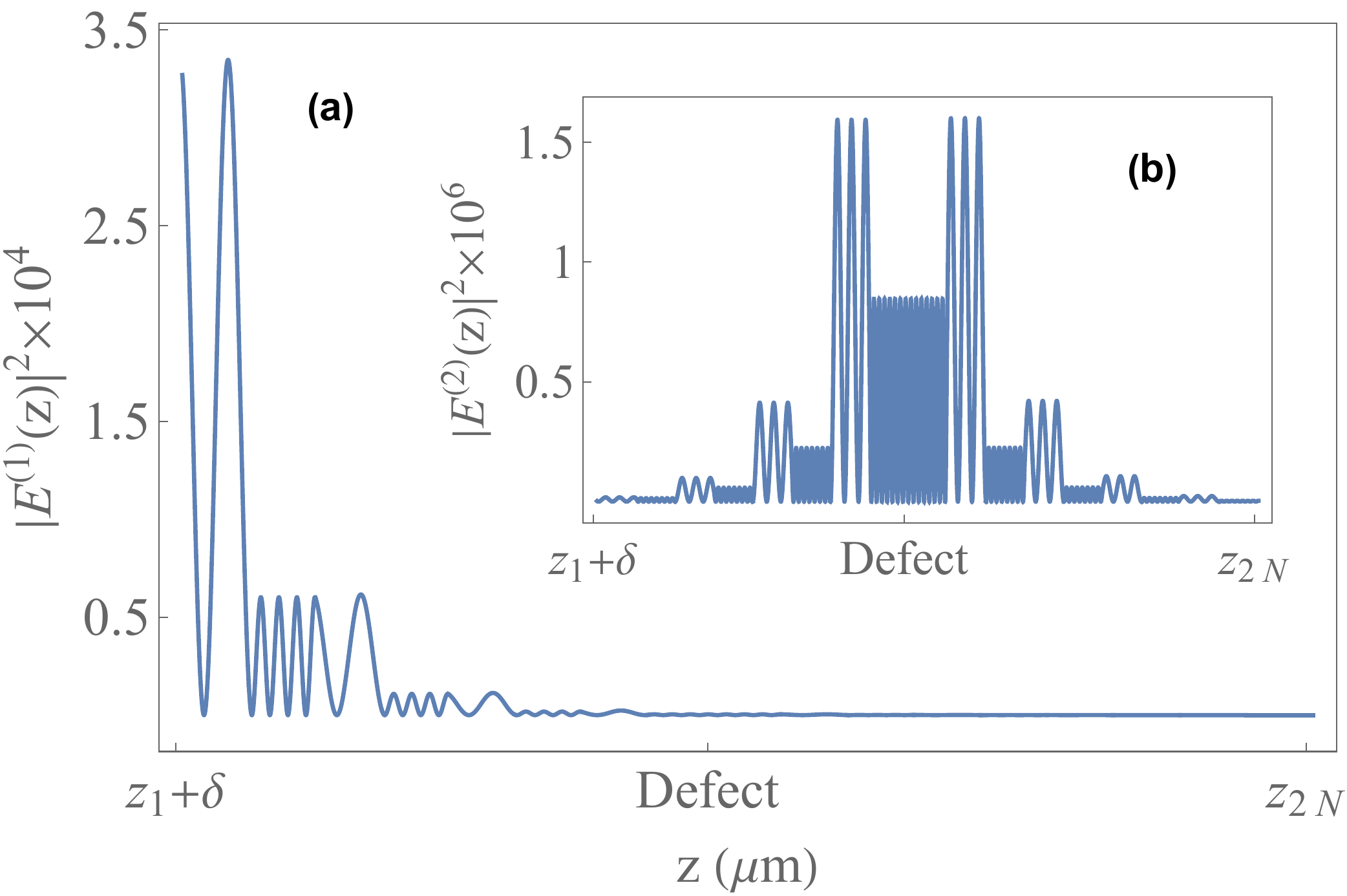}\hspace{0.5cm}
		\includegraphics[scale=.35]{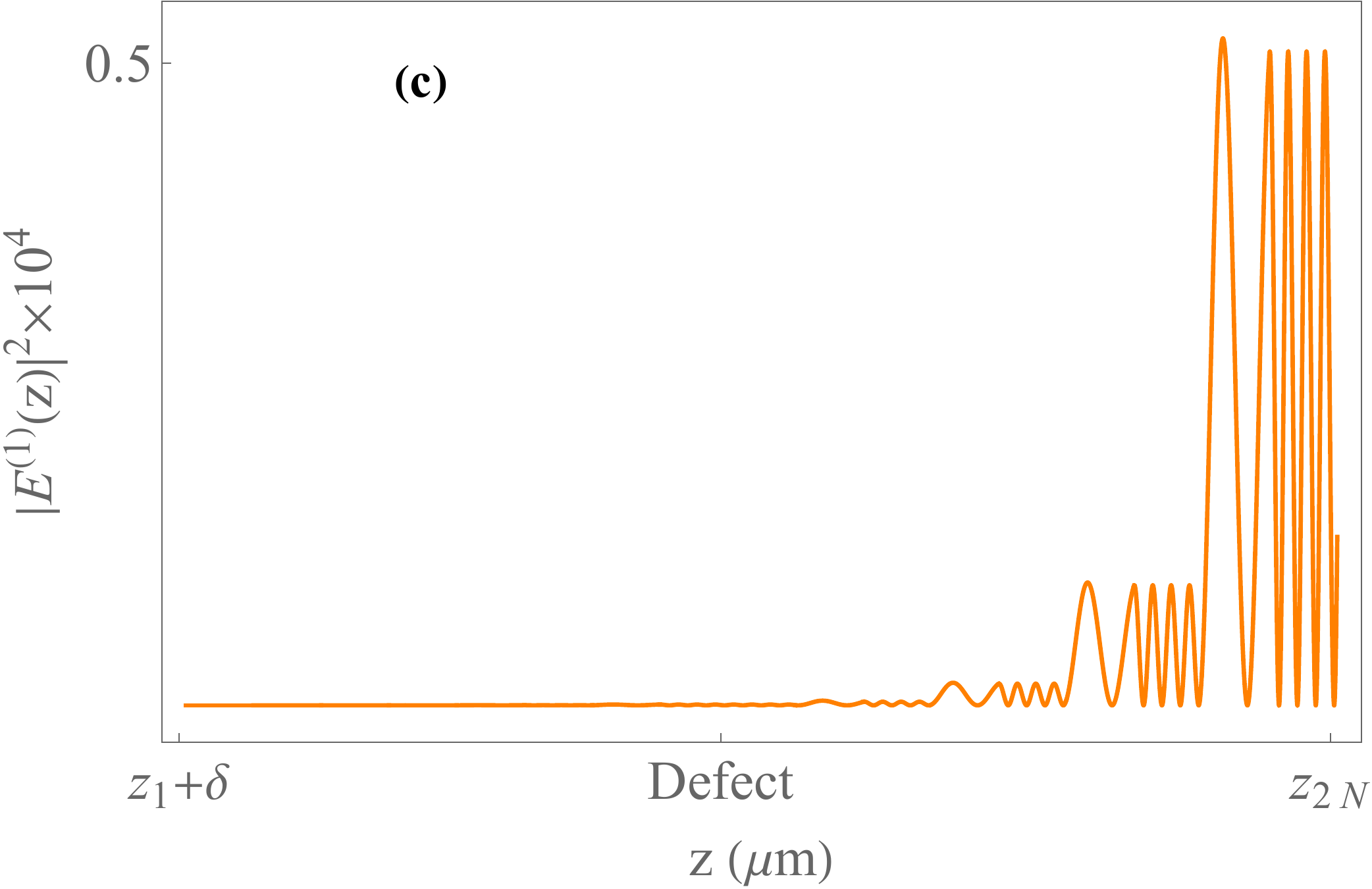}
		\caption{{ Intensity distribution of the left incident FW (a), left incident SHW (b) and right incident FW (c) into linear photonic crystal at $ \lambda_\star=1.736 \mu m$. }}
		\label{distribution}
	\end{center}
\end{figure}
Our strategy for localizing the SHW is to engineer the band structure of the linear lattice $ \mathcal{S}_2 $ by embedding defects in the linear crystal. By taking appropriate parameters of the photonic crystal and its defect, we can get the wave-number $ k_\star:=\frac{\omega_\star}{c} $ for the fundamental wave lies in the stopband, and simultaneously, the wave-number $ 2k_\star $ for the SHW stands on the passband.\\
Allowing that, on the right hand of Eq. \ref{SHG scattering solution}, the transmission coefficient for the SHW, i.e., $ t^{(2)}(2k_\star) $ takes finite value while for the FW, it vanishes. Subsequently, it can be seen easily from Eqs. \ref{SHG scattering solution} and \ref{FW scattering solution} for the $ \omega=\omega_\star $, that the transmitted amplitudes satisfy the following relations
\begin{align}
	&A_+^{(1)}=0, &A_+^{(2)}\notin0.
\end{align}
This solution leads to a localized SHW with frequency $ 2\omega_\star $ in the defect layer of the linear system and exponentially decayed FW \cite{figotin1998localized}. The wave-number lies in the stopband for the right incident wave with frequency $ \omega_\star $, which means the transmission coefficient is zero.By substituting $ t(\omega_\star) $ in Eqs. \ref{R Scattering Eq1} and \ref{R Scattering Eq2}, we have 
\bea
A_0^{(1)}=B_0^{(1)}=0.
\eea
In this case, the photonic crystal totally reflects the wave, and there is no passing FW in the nonlinear medium. In other words, the source term generating a second harmonic is absent. Consequently, the right incident fundamental wave exponentially decays in the linear lattice.\\ In Fig. \ref{fig:setup}, we plot the transmission ad reflection for the linear multilayer slab with a defect layer. In our design, the defective linear crystal consists of $ N=8 $ segments each made of two sections with refractive index $ n_1=1.2 $ (Blue layer) and $ n_2=3.2 $ (Orange layer) and sections take the same thickness, i.e., $ d_1=d_2=1\mu m $. The linear structure becomes defective by making twice the thickness of the second section of the fourth segment $ (d_3=2\mu m) $.\\
The transmission coefficient $ T:=\vert t \vert^2 $ and reflection coefficient $ R:=\vert r \vert^2 $ are given by \ref{Linear T and R}. We plot them in terms of the wavelength in Fig.\ref{fig:setup}. One can see that defect state appears within the photonic bandgap between 0.924-0.821 $ \mu m $ (left diagram) and
2.094–1.570 $ \mu m $ (right diagram). For this defective structure, a fundamental wave with $ \omega_\star=10.783\times 10^2 $Hz is trapped through the crystal while the generated wave with frequency $ 2\omega_\star $ is transmitted. 
In Fig. \ref{fig:transmittion}, we plot the intensity of the transmitted wave  $ \vert E_+^{(1)}\vert^2=\vert A_+^{(1)}\vert^2  $ ($ \vert E_+^{(2)}\vert^2=\vert A_+^{(2)}\vert^2 $) for the FW (SHW) on the right side of the scattering setup. In the calculation, the second-order nonlinear coefficient for the nonlinear slab is $ \chi^{(2)}=100$ pm/V and the refractive index for the FW and SHW is respectively $ \eta_1=3.21 $ and $ \eta_2=3.22 $. Since the efficiency of the nonlinear medium is very low, we assume that a strong light with an amplitude $ \vert E_0\vert=\vert A_-\vert=100V/\mu m $ incident normally on the system. In this case the localization and the transmission of the passing SHW is detectable.
The intensity distribution of the FW and SHW into the linear defective crystal is plotted in Fig. \ref{distribution}. For the left incident wave, the SHW localizes through the defective layer while the FW exponentially decays. We also show the mode distribution of the right coming wave. One can see that the FW exponentially decays in the linear lattice.\\
\begin{figure}
		\includegraphics[scale=.5]{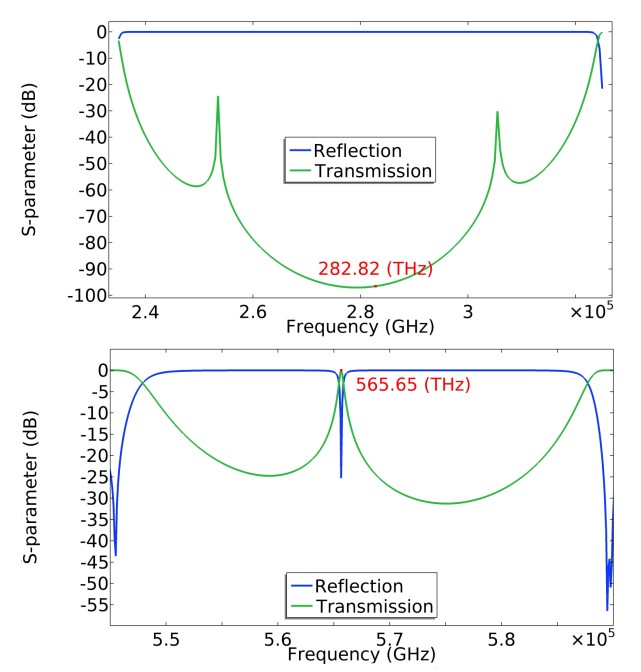}
		\caption{{ Logarithmic diagram of transmission $ T:=\vert t\vert^2 $ and reflection $ R:=\vert r \vert^2 $ spectrum of the defective linear photonic crystal in the numeric simulation. In the upper diagram, the frequency of the fundamental wave ($ \omega_\star=282.82$ Hz ) lies in the bandgap while the second harmonic generated frequency $ 2\omega_\star $ is located on the passband.}}
		\label{RT with defect COMSOL}
\end{figure}

\section{Numeric Simulation}
\begin{figure}[h]
	\includegraphics[scale=0.31]{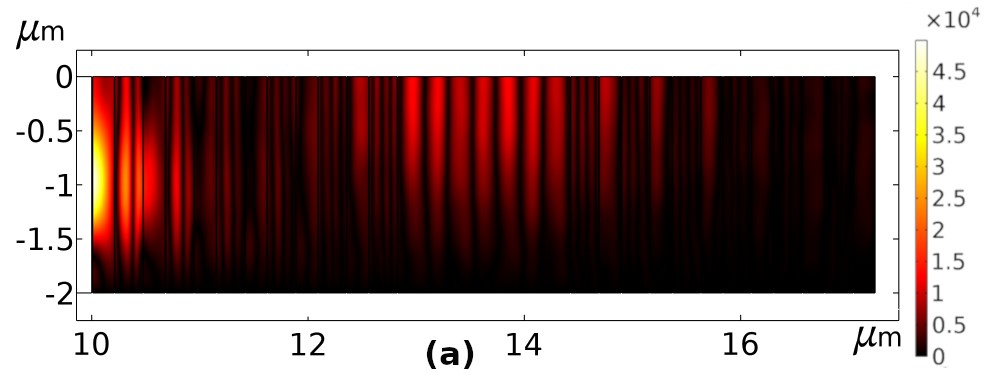}\hspace{0.5cm}
	\includegraphics[scale=0.32]{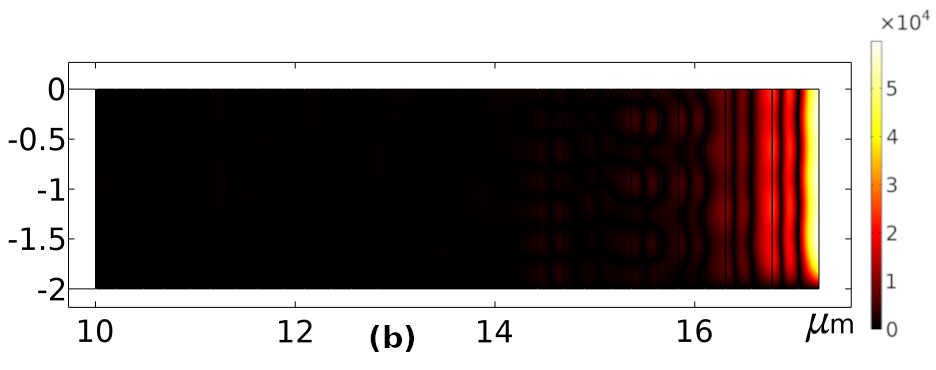}
	\caption{{{In these figures, we demonstrate the mode density
				throughout the linear crystal. For the left incident wave (a), the
				higher intensity on the left side corresponds to the Fundamental
				wave. One can see the localized SHW in the middle of the diagram
				in the defect layer. For the right incident wave (b), the higher intensity on the right side corresponds to the exponentially decaying
				fundamental wave.
	}}}  \label{fig:localisationCOMSOL}
\end{figure}

This section numerically demonstrates second harmonic localization by using finite element method simulations in a time-dependent area. We use COMSOL Multiphysics to perform a time-domain transient simulation of a sinusoidal wave passing through an optical setup which is similar to the one we depict in Fig.\ref{fig:setup}. In our codes, we consider that our linear crystal consists of 12 segments made of two slabs with refractive index $ n_1=1.2 $ and $ n_2=3.2 $. Our linear system has been made defective by adding an extra slab in the middle of the crystal. The optimization of the linear system for finding appropriate bandgap in the way that the generated frequency stands on the passing band determines the length of slabs and defect layer. The optimization defines the length of slabs such that $ d_1=0.2\mu m $, $ d_2=0.26\mu m  $ and $ d_3=1.593 \mu m $. Fig. \ref{RT with defect COMSOL} illustrates the corresponding transmission and reflection amplitudes of the linear crystal in the logarithmic scale.
\begin{figure}[h!]
	\includegraphics[scale=0.30]{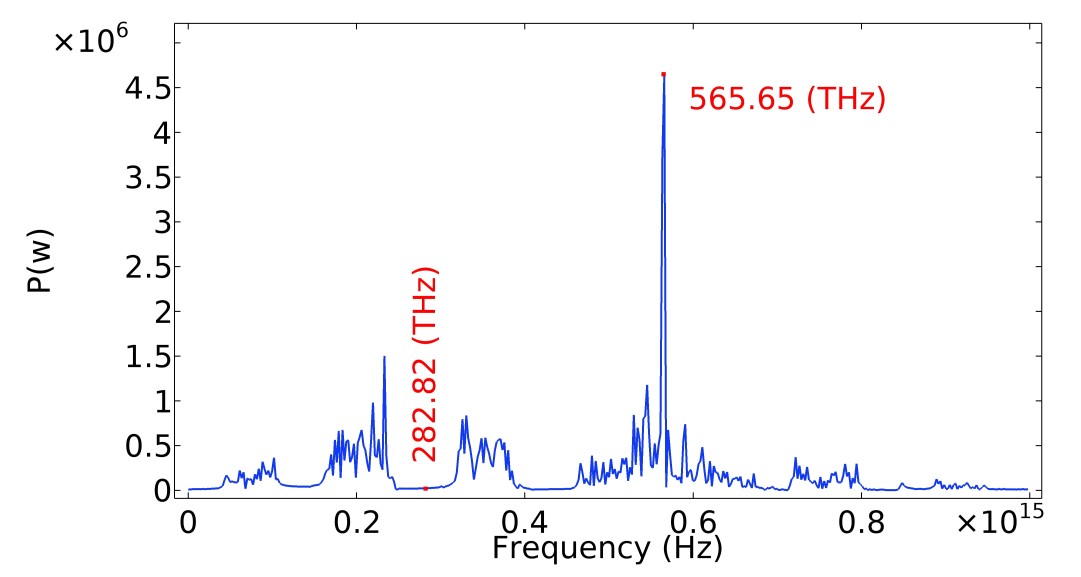}
	\caption{{ This plot shows the contribution of the transmitted wave ($ P_i(\omega):=\vert a_i\vert^2 $) versus frequency derived from the Fourier transform of the time domain transient. The first and second marked point represents $ P_1(\omega_\star) $ and $ P_2(2\omega_\star) $ for the fundamental and the second harmonic}}  \label{transientCOMSOL}
\end{figure}
Regarding these plots, one can see that the asymmetric localization takes place for the left incoming photon with frequency $ \omega_\star=282.82$ Hz. The nonlinearity is enroled in our simulation by considering the coupling between fundamental wave and second harmonic wave via the following polarization
\begin{align}
	&\bP_{1y}=2d_{\text{eff}}E_{2y}E^{*}_{1y},&&\bP_{2y}=d_{\text{eff}}E^2_{1y},
\end{align}\\ 
where we aligine our polarization in $ y $-direction and $ d_{\text{eff}} $ is nonlinear coefficient for the SHG process.
We then probe the transient on the right side of our system and measure the amplitude of time dependant electric field, i.e., $ E_{+y}(t) $. The corresponding contribution of an electric field for different modes in the frequency domain is given by the following Fourier transform
\be
E_{+y}(t)=\int \fa_n(\omega)e^{-in\omega t}dt.
\ee
In Fig. \ref{transientCOMSOL}, we show the mode contribution of the FW at $ \omega_\star=282.82 $ THz and SHW at $ 2\omega_\star=565.65 $ THz. The density of the electromagnetic wave in the linear photonic crystal is depicted in Fig. \ref{fig:localisationCOMSOL}(b) showing that the FW is exponentially decayed in the defective linear system while the SHW is localized in the defect layer. For the right incident wave with the frequency $ \omega_\star $, we find the mode intensity in Fig. \ref{fig:localisationCOMSOL}(c) and show that the FW exponentially decays in the photonic crystal.
\section{Conclusion}
In summary, we suggest an optical system that makes the photon localized when it hits the system from one side. In our method for asymmetric localization, the incident wave is reflected from the optical device while the generated harmonic wave transfers. The system is adjustable in which one can manipulate the linear crystal and its defect to localize higher generated harmonic. The main concern which is issued here is the efficiency of the nonlinear slab. In this paper, we apply the optimization method to increase efficiency to see the effect of localization. We also consider the large value of nonlinear coefficient $ (d_{\text{eff}}) $ in our numeric code to allow for a detectable transient field. The values of physical parameters that we find for refractive index and $ (d_{\text{eff}}) $ can be actualized by applying some method such as quasi-phase-matching.
\begin{acknowledgments}
We acknowledge the support by the Army Research Office Grant No. W911NF-20-1-0276 and NSF Grant No. PHY-2012172. The views and conclusions contained in this document are those of the authors and should not be interpreted as representing the official policies, either expressed or implied, of the Army Research Office or the U.S. Government. The U.S. Government is authorized to reproduce and distribute reprints for Government purposes notwithstanding any copyright notation herein. 
\end{acknowledgments}

\appendix

\section{}
In this section, we construct the nonlinear transfer matrix of a nonlinear layer under the non-depletion signal approximation\cite{li2007second,li2015application}. In terms of the definition of transfer matrix \cite{mostafazadeh2020transfer} and continuity condition for electric field $ \vec{E} $ and magnetic field $ \vec{H} $ at $ z=z_0,z_1 $ \cite{jackson1999classical}, the transfer matrix $\bM_2$ of slab $\cS_1$ for second harmonic generation can be given by the following relation
\bea
&\left(\begin{array}{cc}
	A_{0}^{(2)}   \\
	B_{0}^{(2)}
\end{array}\right)=\cM_{L}^{(2)}\left(\begin{array}{cc}
	0    \\
	B_{-}^{(2)}
\end{array}\right)\nonumber\\
&+\chi^{(2)} \cM_{N}^{(2)}(A_{-}^{(1)},B_{-}^{(1)})\left(\begin{array}{cc}
	A_{-}^{(1)}    \\
	B_{-}^{(1)}
\end{array}\right).
\label{ANp1BNp1}
\eea 
The first term on the right side presents the linear transfer matrix which is generated by the homogenous solution of Eq. \ref{nceq2} and relates the free-wave amplitude of the second harmonic field on te both sides of the nonlinear medium. It is given by the following relation
\be
\cM_{L}^{(2)}=\dfrac{1}{\eta_2}\textbf{Q}_{-1}^{(2)}\textbf{K}_{+}^{(2)}\textbf{P}_{+1}^{(2)}\textbf{P}_{-0}^{(2)}\textbf{K}_{-}^{(2)}\textbf{Q}_{+0}^{(2)}.
\ee
The related matrices are defined as
\bea
&&	\bQ_{\pm j}^{(i)}=\left( \begin{array}{cc}
	e^{\pm ik_iz_j}	&0  \\
	0& e^{\mp ik_i
		z_j}
\end{array}\right),\bP_{\pm j}^{(i)}=\left( \begin{array}{cc}
	e^{\pm i\tilde{k}_iz_j}	&0  \\
	0& e^{\mp i\tilde{k}_iz_j}
\end{array}\right),\nonumber\\
\nonumber\\
&&~~~~~~~~~~~~\bK_\pm^{(j)}=\frac{1}{2\eta_j}\left( \begin{array}{cc}
	\eta_{j}+1	&\pm\eta_{j}\mp 1  \\
	\pm\eta_{j}\mp 1& \eta_{j}+1
\end{array}\right),	
\eea
and $ \tilde{k}_j=\eta_j k_j $.
The second part in Eq. \ref{nceq2} denotes the nonlinear transfer matrix which relate the bound-wave amplitudes of the second harmonic field created by the fundamental wave. We find the nonlinear transfer matrix as:
\bea
&&\cM_{N}^{(2)}=\dfrac{\cG_1}{2\tilde{k}_2}\bQ_{-1}^{(2)}[\bK\textbf{O}_1-\tilde{\bK}\textbf{O}_0]\bN_1\bP_{-0}^{(1)}\bK_+^{(1)}\bQ_{+}^{(1)}\nonumber\\
&&+\cG_2\bQ_{-1}^{(2)}\bN_2\bP_{-0}^{(1)}\bK_+^{(1)}\bQ_{+}^{(1)}
\eea
where
\bea
&&	\tilde{\bK}=\left( \begin{array}{cc}
	\tilde{k}_2+2\tilde{k}_1	&\tilde{k}_2-2\tilde{k}_1  \\
	\tilde{k}_2+2\tilde{k}_1& \tilde{k}_2-2\tilde{k}_1
\end{array}\right),\bK=\left( \begin{array}{cc}
	k_2+2\tilde{k}_1	&k_2-2\tilde{k}_1  \\
	k_2+2\tilde{k}_1& k_2-2\tilde{k}_1
\end{array}\right),\nonumber\\
\nonumber\\
&&~~~~~~~~~~~~~\textbf{O}_j=\left( \begin{array}{cc}
	e^{2i\tilde{k}_1z_j}	&0  \\
	0& e^{-2i\tilde{k}_2z_j}
\end{array}\right),	
\eea
and, we intruduce amplitude dependant matrices
\bea
&\bN_1=\left( \begin{array}{cc}
	C^{(1)}	&0  \\
	0& D^{(1)}
\end{array}\right),&\bN_2=\left( \begin{array}{cc}
	0	&C^{(1)}  \\
	D^{(1)}& 0
\end{array}\right).
\eea
The amplitudes $ C^{(1)} $ and $ D^{(1)} $ can be defined from the FW such as
\be
\left(\begin{array}{cc}
	C^{(1)}   \\
	D^{(1)}
\end{array}\right)=\bP_{-0}^{(1)}\bK_+^{(1)}\bQ_{+0}^{(1)}\left(\begin{array}{cc}
	A_{-}^{(1)}    \\
	B_{-}^{(1)}
\end{array}\right).
\label{C1D1}
\ee

\end{document}